\documentclass[journal=cmatex,manuscript=article]{achemso}
\setkeys{acs}{articletitle=true}

\usepackage[version=3]{mhchem} 
\usepackage[T1]{fontenc}       
\usepackage{graphicx}
\usepackage{subcaption}
\usepackage{booktabs}
\usepackage{color}
\usepackage{multirow, array}

\author{Davide Stefani}
\altaffiliation{These two authors contributed equally}
\affiliation[TU Delft]
{Kavli Institute of Nanoscience, Delft University of Technology, Lorentzweg 1, 2600 GA, The Netherlands}
\author{Cristian A. Guti{\'e}rrez-Cer{\'o}n}
\altaffiliation{These two authors contributed equally}
\affiliation[USACH]
{Department of Physics, Faculty of Physical and Mathematical Sciences, University of Chile, Av. Blanco Encalada 2008, 8330015 Santiago, Chile}
\author{Daniel Aravena}
\affiliation[USACH]
{Department of Material Chemistry, Faculty of Chemistry and Biology, University of Santiago de Chile, Casilla 40, Correo 33, 9170022 Santiago, Chile}
\author{Jacqueline Labra-Mu\~{n}oz}
\affiliation[USACH]
{Department of Electrical Engineering, Faculty of Physical and Mathematical Sciences, University of Chile, Av. Blanco Encalada 2008, 8330015 Santiago, Chile}
\author{Catalina Suarez}
\author{Shuming Liu}
\affiliation[UTEP]
{Department of Chemistry, University of Texas, 500 West University Avenue, El Paso, Texas 79968, United States}
\altaffiliation{Current address: Hebei Sheng, Handan Shi Xueyuanbeilu 530, Chemistry Department, P. R. China 056005}
\author{Monica Soler}
\affiliation[USACH]
{Department of Material Science, Faculty of Physical and Mathematical Science, University of Chile, Av. Beauchef 851, 8330015 Santiago, Chile}
\author{Luis Echegoyen}
\affiliation[UTEP]
{Department of Chemistry, University of Texas, 500 West University Avenue, El Paso, Texas 79968, United States}
\author{Herre S.J. van der Zant}
\affiliation[TU Delft]
{Kavli Institute of Nanoscience, Delft University of Technology, Lorentzweg 1, 2600 GA, The Netherlands}
\author{Diana Duli\'{c}}
\affiliation[USACH]
{Department of Physics, Faculty of Physical and Mathematical Sciences, University of Chile, Av. Blanco Encalada 2008, 8330015 Santiago, Chile}
\email{ddulic@ing.uchile.cl}
\phone{+56 229784518}

\title[]
  {Charge transport through a single molecule of \textit{trans}-1-\textit{bis}-diazofluorene [60]fullerene}

\begin{tocentry}
	\centering
	\includegraphics[height = 3.6cm]{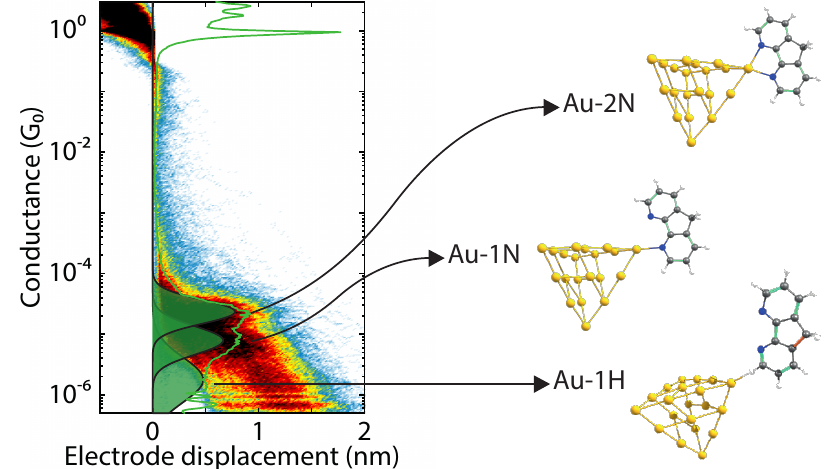}
\end{tocentry}

\begin{document}

\begin{abstract}
Fullerenes have attracted interest for their possible applications in various electronic, biological, and optoelectronic devices.
However, for efficient use in such devices, a suitable anchoring group has to be employed that forms well-defined and stable contacts with the electrodes.
In this work, we propose a novel fullerene tetramalonate derivate functionalized with \textit{trans}-1 4,5-diazafluorene anchoring groups.
The conductance of single-molecule junctions, investigated in two different setups with the mechanically controlled break junction technique, reveals the formation of molecular junctions at three conductance levels.
We attribute the conductance peaks to three binding modes of the anchoring groups to the gold electrodes.
Density functional theory calculations confirm the existence of multiple binding configurations and calculated transmission functions are consistent with experimentally determined conductance values.
\end{abstract}

\section{Introduction}
For many years a wide variety of fullerene derivatives has been reported and their potential applications as electronic, magnetic, catalytic, biological and optical materials have been explored \cite{intro1,intro2,intro3,intro4,intro5,intro6,intro7}.
In particular, C$_{60}$ and its derivatives are ideal candidates for molecular-based devices because of their interesting and unique electronic properties \cite{nano,natmat}.
One of the long-standing challenges in obtaining functional molecular electronic devices is the transport of electrons through single molecules \cite{Xiang2016,Jeong2017}; i.e. it is necessary to guarantee molecular conduction through a single molecule rather than through a group of molecules \cite{wires}.
The electron transport properties of single molecules can then be controlled electrically, magnetically, optically, mechanically or chemically to obtain the desired functionality \cite{natnano}.
However, in the case of fullerene derivatives, the groups used to functionalize the carbon cage may also modify the electronic properties of the molecule \cite{Peyghan2013}.
In addition, the molecule needs to possess anchoring groups that will form well defined and stable molecular junctions with the electrodes \cite{Zotti2010revealing,Kristensen2008, Kaliginedi2014}.
Therefore, the orientation and nature of the anchoring groups will affect the charge transport properties of the molecule.
Finding a suitable anchoring group to connect C$_{60}$ electrically would be an important step towards more interesting derivatives in the future.

Charge transport in pristine C$_{60}$ has been studied for a long time using both scanning tunnelling microscope break junction (STM-BJ) \cite{joachim1995electronic,bohler2007conductance} and mechanically controlled break junction (MCBJ) \cite{MCBJKondo,kiguchi2008conductance} techniques and conductance values around 0.1 G$_{0}$ have been reported; here G$_{0}$ is the quantum of conductance equaling $2e^2/h$ = 77.48 $\mu$S \cite{Ohnishi1998,Landauer1970}.
Taking advantage of its affinity for metals, C. Martin et al. \cite{martin2008fullerene} studied a `dumbbell' benzene-difullerene (BDC60) derivate in which C$_{60}$, at both ends of the molecule, acts as the anchoring group.
The single peak of conductance found at $3 \cdot 10^{-4}$ G$_{0}$ for this derivative was attributed to a transport-limiting barrier created by the nitrogen atoms of the pyrrolidine rings that are connected to the benzene backbone \cite{lortscher2013bonding}.
An analogous work with a dumbbell fullerene molecule reported a conductance peak around $1 \cdot 10^{-4}$ G$_{0}$ confirming the low conductance values for functionalized fullerenes \cite{Leary2011}.
Additionally, 
a more recent study of a dumbbell fullerene derivative shows two different electronic transport configurations, one assigned to the transport through the molecular bridge and the second, at higher conductance values, ascribed to a single C$_{60}$ anchoring group trapped between the two adjacent electrodes \cite{Moreno2015}.
These two peaks differ by two orders of magnitude ($10^{-1}$ G$_{0}$ and $10^{-3}$ G$_{0}$), which is relatively consistent with the previous reported values for pristine C$_{60}$ fullerenes \cite{joachim1995electronic,bohler2007conductance,MCBJKondo,kiguchi2008conductance} and dumbbell C$_{60}$ fullerene derivatives \cite{martin2008fullerene,lortscher2013bonding,Leary2011,Ullmann2015single}.
We can also find theoretical studies that support these results \cite{Bilan2012theoretical,Geranton2013transport} and experimental studies of other C$_{60}$ fullerene derivatives, i.e. amino-terminated derivatives \cite{Morita2008reduction}.

\begin{figure}[p]
    \centering
    \begin{subfigure}[b]{0.65\linewidth}
        \includegraphics[width=\linewidth]{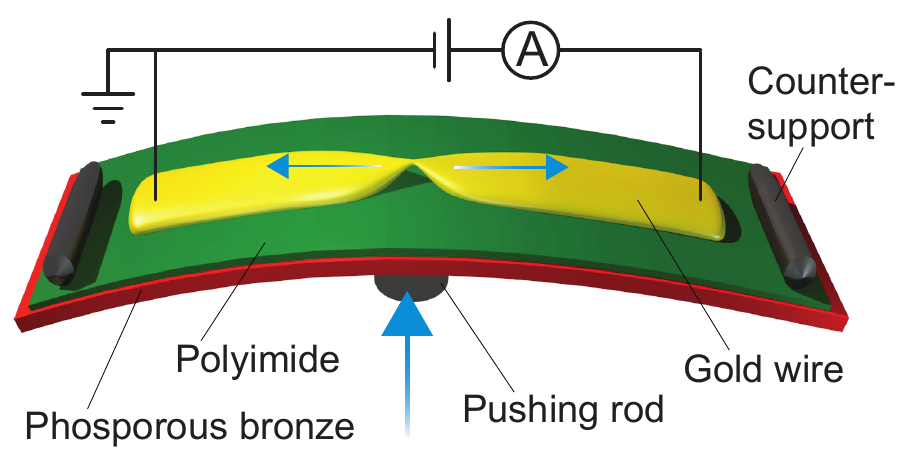}
    \end{subfigure}
    \hfill
    ~ 
    \begin{subfigure}[b]{0.3\linewidth}
        \includegraphics[width=0.75\linewidth]{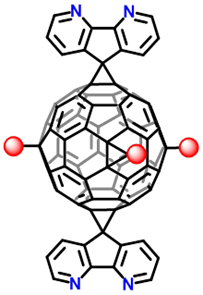}
    \end{subfigure}
    \hfill
    \caption{a) Schematics of the MCBJ devices used in the experiment. \cite{martin2008lithographic} b) Chemical structure of the \textit{trans}-1-bis-diazofluorene adduct of C$_{60}$-tetramalonate (C$_{60}$-daf). Malonates are designated with red balls representing the \ce{C(COOC2H5)2} groups.}
	 \label{fig:schematics}
\end{figure}

In this work, we study the molecular conductance through single-molecule junctions of a C$_{60}$ fullerene all-equatorial tetramalonate derivate functionalized with \textit{trans}-1 4,5-diazofluorene anchoring groups \cite{peng2013high} (see structure in Fig. \ref{fig:schematics}b and hereafter denoted as C$_{60}$-daf for convenience).
The diazafluorene anchoring groups are perfectly perpendicularly oriented with respect to the fullerene surface, and linearly disposed with respect to each other. 
Four equatorial diethyl malonate groups were added to the C$_{60}$ to avoid direct interactions between the gold electrodes and the fullerene cage.
The ethyl esters in these groups have high contact resistance and low binding energy to gold, which effectively renders them bulky groups that block access to the C$_{60}$ cage\cite{Chen2006Effect}.
This aspect and the evidence of direct coordination between the diazafluorene groups and many metal ions \cite{annibale2016coordination,peng2013high} suggest that the formation of a molecular junction occurs only through a linear arrangement of diazafluorene groups and the gold electrodes. Such a junction structure allows the study of charge transport through the carbon cage of  the fullerene derivative.
The computational simulations performed indeed show preferential binding of the nitrogen atoms to the gold leads and that the highest occupied molecular orbital (HOMO) is sufficiently well communicated to allow the electronic transfer.

Because one of the biggest challenges in molecular electronics field is to have comparable and reproducible results between different techniques, instruments, or data sets, we have performed our experiments in two identical MCBJ setups located in two different laboratories in Delft and Santiago de Chile.

To the best of our knowledge, the diazafluorene group has never been tested as an anchoring group before \cite{Leary2015Incorporating,Metzger2015Unimolecular}.
In addition, charge-transport studies using a C$_{60}$ derivative with \textit{trans}-1 terminal groups to form molecular junctions with gold electrodes have not been reported.

\section{Results and discussion}
The conductance of C$_{60}$-daf (Fig. \ref{fig:schematics}b) was measured in air at room temperature.
Both MCBJ setups (the setups are copies and made in Delft/Leiden) and the measuring technique have been extensively described elsewhere\cite{martin2008lithographic,martin2011versatile}.
To fabricate the MCBJ devices (Fig. \ref{fig:schematics}a), e-beam lithography is used to deposit a gold wire with a thin (<100 nm wide) constriction on a bendable substrate. The substrate is then bent by driving a pushing rod onto its middle part while keeping its edges clamped.
This causes the gold wire to stretch until a single gold atom connects the two extremities.
Further bending results in a breaking of the wire forming two sharp electrodes.
The single-gold atom termination can be observed in a conductance vs. electrode displacement trace (breaking trace) as the formation of a plateau around 1 G$_{0}$.
A sudden drop in conductance to about $10^{-3}$ G$_{0}$ signals the rupture of the gold wire.
This point is defined as the zero displacement ($d=0$) in a breaking trace.
After the initial opening of the junction, the electrodes are moved apart and the conductance is recorded until the noise level is reached.
When this sequence is finished, the electrodes are fused back together until the conductance is above 30 G$_{0}$ and the whole process is repeated.

We start each experiment by characterizing the bare device.
We apply a bias voltage of 0.1 V to the gold wire and measure the current passing through it while repeatedly opening and closing the junction.
A device is used for molecule measurements only if it shows just vacuum tunneling and a clear single-gold atom, 1 G$_{0}$ plateau\cite{frisenda2013statistical}.
An $\sim$30 $\mu$M solution with the molecules under investigation is prepared by dissolving the starting compound in dichloromethane.
Two 2 $\mu$L droplets of the solution are subsequently drop-cast on the freshly characterized device.

\begin{figure}[p]
    \centering
      \includegraphics[width=0.9\linewidth]{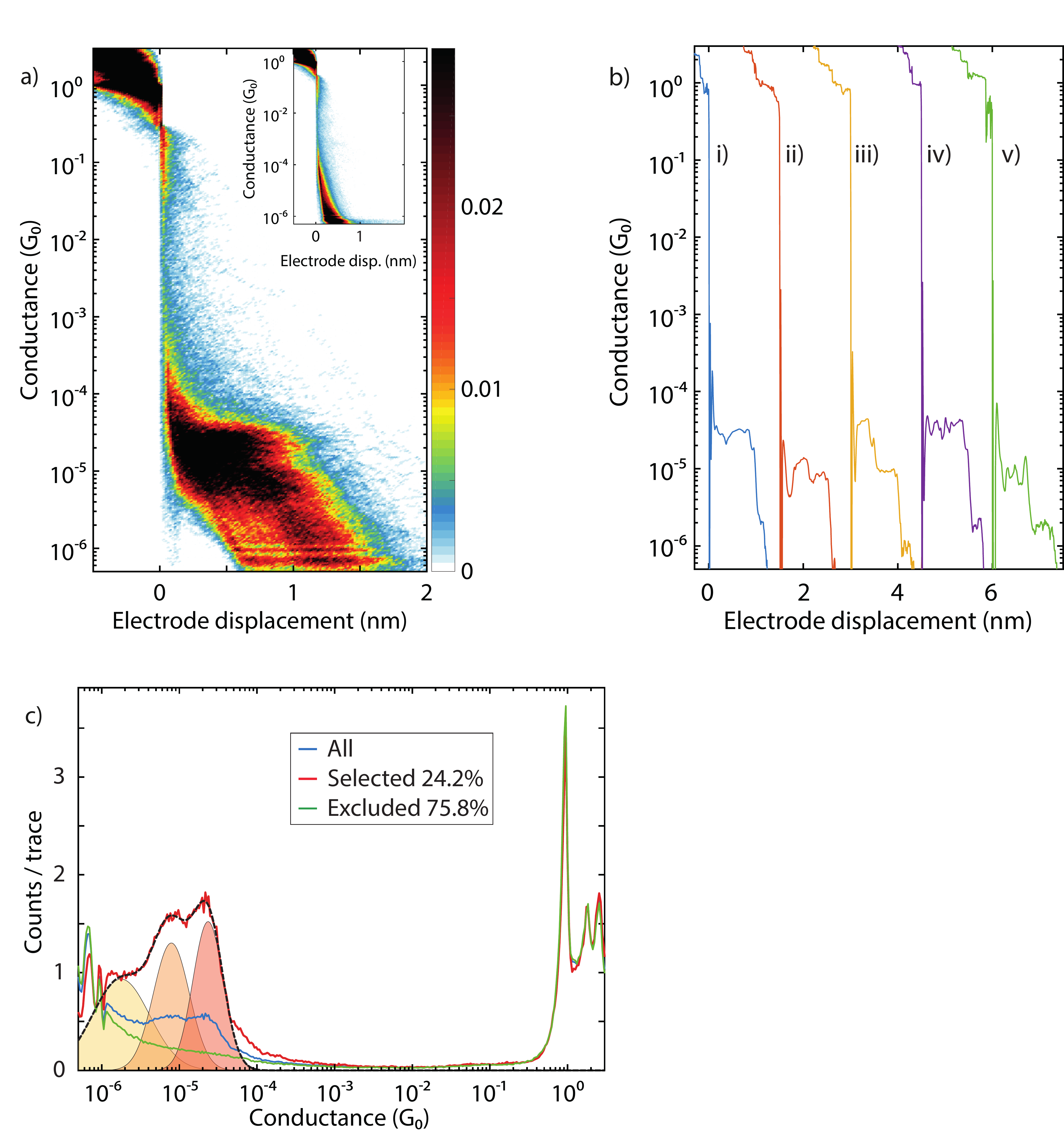}\\
    \caption{a) Two-dimensional conductance histogram built from a selection of the 10.000 consecutive breaking traces recorded after drop-casting the solution containing C$_{60}$-daf. Traces that had 1.2 times the average amount of counts between $1 \cdot 10^{-4}$ and $1 \cdot 10^{-6}$ G$_0$ were selected (24.2\% of the total). The inset shows the traces that were not selected  (i.e., that have less than 1.2 times the average count in the region under consideration, 75.8\%). The breaking traces have been logarithmically binned in the conductance axis with 49 bins/decade and with 58.4 bins/nm in the displacement axis. The applied bias voltage is 0.1 V and the electrode speed is 4.0 nm/s.  b) Individual breaking traces in the presence of a C$_{60}$-daf molecule. Traces are offset along the x-axis by 1.5 nm for clarity. c) Normalized one-dimensional histograms obtained by integrating the breaking traces along the displacement axis. The red line represents the histogram obtained for the selected traces, the green line from the excluded traces, and the blue line shows the histogram of the whole data set. The dashed black line represents the log-normal fit to the selected histogram (red line). Experiment conducted in Delft, The Netherlands.}
    \label{fig:MCBJDelft}
\end{figure}

A data set is composed of thousands of consecutive breaking traces from individual junctions recorded with the same settings and is used to construct a 2D conductance histogram (conductance vs. electrode displacement)\cite{frisenda2013statistical}.
By integrating over the displacement, a one-dimensional histogram is obtained, from which the most probable junction conductance is usually estimated.
To facilitate the identification of molecular traces, a  home-made MatLab program is used to select traces that have high counts in the conductance region of interest.
The filtering method is based on the fact that if a molecule is not trapped in the junction the conductance decreases exponentially; the corresponding breaking trace therefore does not display many counts in the high-conductance region.
The filtering procedure can also be used to estimate the percentage of junctions that contain a molecule.
To construct the 2D histogram in Fig. \ref{fig:MCBJDelft}a, we used the following criterion: traces that have 1.2 times the average amount of counts in the $1 \cdot 10^{-4}$ to $1 \cdot 10^{-6}$ G$_{0}$ region are selected.
The inset in the same figure displays the traces that did not satisfy the requirements and were therefore excluded from the selection.
The sum of these two histograms thus constitutes the complete data set, which can be found in the Supporting Information (Fig. S1).
We have furthermore verified that when adjusting the filtering criteria the main conclusions of the paper do not change (see Supporting Information Fig. S4-S5).

The 2D histogram in Fig. \ref{fig:MCBJDelft}a shows a high-count region around $10^{-5}$ G$_{0}$, which extends up to 1.5-2 nm.
The counts are concentrated mostly around two values and a log-normal fit of the one-dimensional histogram (Fig. \ref{fig:MCBJDelft}c) indicates that the corresponding most probable conductance values are $2.3 \cdot 10^{-5}$ (peak A) and $7.9 \cdot 10^{-6}$ G$_{0}$ (peak B).
The individual breaking traces show that the plateaus can (traces i-ii in Fig. \ref{fig:MCBJDelft}b) but do not always appear together (trace iii).
Some breaking traces also show a third plateau around $2 \cdot 10^{-6}$ G$_{0}$ (traces iv-v in Fig. \ref{fig:MCBJDelft}b), but their appearance is not as frequent.
Including this third peak in the fitting increases the accuracy of the fit and yields a conductance value of $1.8 \cdot 10^{-6}$ for peak C.

\begin{figure}[p]
    \centering
 		\includegraphics[width=\linewidth]{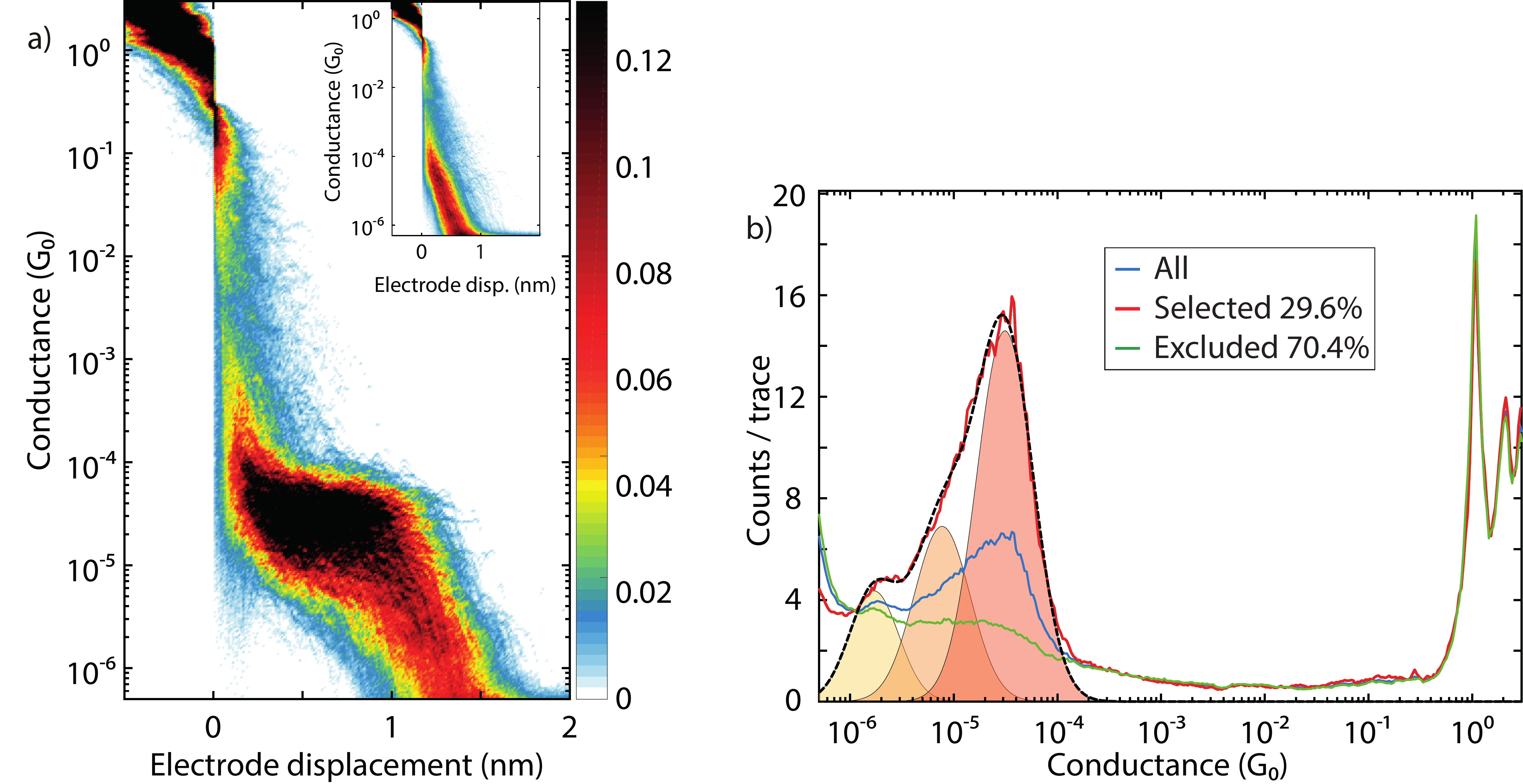}\\
    \caption{a) Two-dimensional conductance histogram built from a selection of the 5.000 consecutive breaking traces recorded after drop-casting the solution containing C$_{60}$-daf. Traces that had 1.2 times the average amount of counts between $1 \cdot 10^{-4}$ and $1 \cdot 10^{-6}$ G$_0$ were selected (29.6\% of the total). The inset shows the traces that were not selected  (i.e. that have less than 1.2 times the average count in said region, 70.4\%). The breaking traces have been logarithmically binned in the conductance axis with 49 bins/decade and with 88.4 bins/nm in the displacement axis. The applied bias voltage is 0.1 V and the electrode speed is 6.0 nm/s.  b) Normalized one-dimensional histograms obtained by integrating the breaking traces along the displacement axis. The red line represents the histogram obtained for the selected traces, the green line from the excluded traces, and the blue line shows the histogram of the whole data set. The dashed black line represents the log-normal fit to the selected histogram (red line). Experiment conducted in Santiago de Chile, Chile.}
    \label{fig:MCBJChile}
\end{figure}

The experiment has been repeated in Santiago de Chile by the group of D. Duli\'{c}.
Fig. \ref{fig:MCBJChile}a shows the two-dimensional histogram obtained from a selection of the 5.000 traces collected (same criterion as used in Fig. \ref{fig:MCBJDelft}a; a histogram made from the rest of the traces is shown in the inset).
The same bias voltage of 0.1 V as in Delft was used, but the electrode speed was one and a half times higher in this case.
The plot confirms the same high-count region near $10^{-5}$ G$_{0}$, displaying a striking resemblance with the one measured in Delft.
In addition, the histogram measured at Santiago de Chile more clearly shows the high counts in the low-conductance region, centred around $2 \cdot 10^{-6}$ G$_{0}$ and extending to lengths of 1.5 nm.
The one-dimensional histogram (Fig. \ref{fig:MCBJChile}b) highlights this area with a peak in conductance, from which the most probable conductance value of $1.7 \cdot 10^{-6}$ G$_{0}$ (peak C) is obtained.
The log-normal fit of the higher conductance region yields $3.1 \cdot 10^{-5}$ (peak A) and $7.7 \cdot 10^{-6}$ G$_{0}$ (peak B), values which are close to those found in the measurements performed in Delft.

From the conductance histograms in Fig. \ref{fig:MCBJDelft}-\ref{fig:MCBJChile} it is evident that peak A represents the most probable formation of molecular junctions in both measurements.
The measurements obtained in Delft show the presence of peak B more clearly, whereas peak C is more pronounced in the measurements from Santiago de Chile.
A comparison between the fit results obtained from fitting the data set acquired in Delft and in Santiago de Chile can be seen in Tab. \ref{tbl:FitResult}.
The values obtained from two other samples are also included within parenthesis, with the corresponding conductance histograms shown in the Supporting Information (Fig. S2-S3).
The conductance values for the three peaks obtained in Delft and in Santiago de Chile match very well.
The width ($w$) of peaks A and B is very similar in both cases, whereas that of peak C is different.
We attribute this difference to the difficulty of fitting peak C in the sample obtained in Delft.
A more refined data selection shown in the Supporting Information shows an improved fitting of this peak (Fig. S4).
Furthermore, peak A is more prominent in the measurements from Santiago de Chile, while its importance is less striking in those from Delft.
This is true for all samples except one presented in the Supporting Information (Fig. S3).
Peak C is the smallest one in all measurements.

\begin{table}[p]
  \begin{tabular}{c ccc ccc}
    \toprule
    	&	\multicolumn{3}{c}{Delft} & \multicolumn{3}{c}{Santiago de Chile}\\
	 \cmidrule(lr){2-4}    \cmidrule(lr){5-7}
    Peak	&  $\mu$ ($10^{-5}$ G$_0$) & $p$  & $w$ &  $\mu$ ($10^{-5}$ G$_0$) & $p$  & $w$\\
	 \cmidrule(lr){1-7}
    A	& 2.3 (2.3) 	& -4.7 (-4.6)	& 0.6 (0.5) & 3.1 (2.8)		& -4.5 (-4.6)	& 0.6 (0.6)\\
    B	& 0.79 (0.87)	& -5.2 (-5.1)	& 0.5 (0.7) & 0.77 (0.82)	& -5.1 (-5.1)	& 0.6 (1.3)\\
    C	& 0.18 (0.12)	& -5.8 (-5.9)	& 0.9 (1.7) & 0.17 (0.16)	& -5.8 (-5.8)	& 0.6 (0.4)\\
	\bottomrule
  \end{tabular}
  \caption{Comparison of the fitting parameters obtained from the measurements in Delft and Santiago de Chile.
  The $\log_{10}$ of the peaks found in the one-dimensional conductance histogram after data filtering were fitted to a triple-log normal ($y = h \, e^{- \big(\tfrac{\log_{10} \left(G/G_0\right) \, p} {w} \big) ^2}$), where $p$ is the peak maximum expressed in $\log_{10}\left(G/G_0\right)$, $\mu$ is the corresponding conductance in G$_{0}$, $w$ is the full-width-half-maximum of the peak expressed in $\log_{10}\left(G/G_0\right)$, and $h$ is the peak height (not displayed in the table).
  The results obtained from two other samples are included within parenthesis.
  The conductance histograms of these samples are shown in the Supporting Information.}
  \label{tbl:FitResult}
\end{table}

Recently, Seth \latin{et al.} \cite{seth2017conductance} studied the conductance through a bis-terpyridine derivate with the MCBJ technique, finding that multiple configurations that display short plateaus are accessible for charge-transport with the conductance spanning several orders of magnitude.
Although we could expect a similar behaviour from the diazafluorene group, the number of accessible conductance configurations should be less because of the reduced number of nitrogen atoms.
Moreover, the C$_{60}$ and malonate groups contribute to the rigidity of the molecule, resulting in more stable conductance plateaus, as clearly shown in the data presented in Fig. \ref{fig:MCBJDelft}-\ref{fig:MCBJChile}.
We tested the behaviour of this molecule in our setup and to be able to do a more accurate comparison with these previous result we measured its single molecule conductance. In this case we did not observe well defined conductance values (see Supporting Information, Fig. S6), again suggesting that the presence of the C$_{60}$ molecular bridge plays an important role in obtaining well defined molecular junctions.
Moreover, the additional nitrogen binding site could provide configurations with more similar conductance to each other, thus hindering their identification.

Although the measurements cannot directly discard the involvement of the aromatic rings in the diazafluorene units in the charge transport, we would expect a more continuous decrease in conductance if they were involved, instead of the step-like plateaus observed in the breaking traces.
The more continuous decay would in turn result in a broader distribution of the conductance values instead of the agglomeration around only three separate values, each with a variance comparable to that of other experiments that employ amino anchoring groups \cite{Frisenda2015,Hong2012single}.

\begin{figure}[H]
	\includegraphics[width=\linewidth]{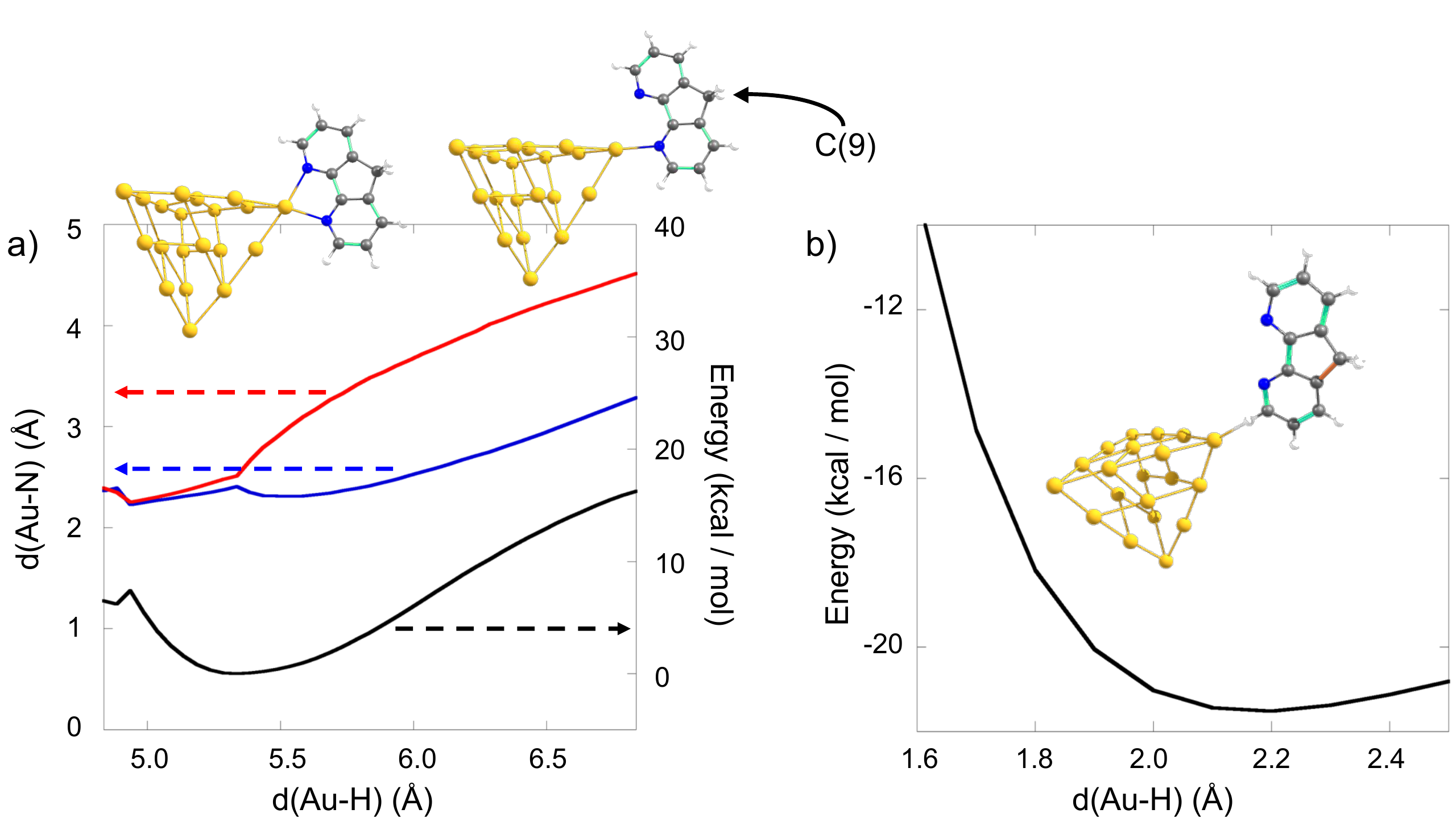}
   \caption{a) Au$-$N distance (blue and red lines) and total energy (black line) at each step of the surface scan of the diazafluorene-Au$_{20}$ model. Snapshots of a two- and one- N coordinated geometry are depicted in the graphic. b) Potential energy surface for the Au$-$H interaction.}
    \label{fig:PotentialEnergy}
\end{figure}

To gain additional insight on the binding configurations of C$_{60}$-daf, \latin{ab-initio} density functional theory calculations have been performed.
The transmission function $T$($E$) at zero bias was calculated using the B3LYP functional (see Computational Details section for a detailed description of the computational methodology).
To consider the different binding modes of the diazafluorene group, a geometry optimization of this anchoring group and a Au$_{20}$ cluster was performed.
The lowest energy conformation corresponds to a geometry in which the Au$_{20}$ cluster is coordinated to the two N-donor atoms with a distance of ca. 2.45 \AA.
A relaxed surface scan was then run, elongating the distance between the top Au atom and the central C atom  of the diazafluorene moiety (position 9 of the fluorene moiety, see Fig. \ref{fig:PotentialEnergy}).
As the distance constraint is not defined with respect to the Au$-$N distance, the nitrogen atoms are free to accommodate their position with respect to the gold tip during the scan. 

A clear rupture of one of the Au$-$N bonds occurs at a Au$-$C(9) distance of 5.35 \AA, when one of the Au$-$N bonds remains short (blue line) while the other is markedly longer (red line) (Fig. \ref{fig:PotentialEnergy}a).
Considering the orientation change of the diazafluorene group during the breaking of the Au$-$N contacts, it is also convenient to explore the possibility of a van der Waals contact between the Au and the H atoms in the diazafluorene.
Such interactions can yield additional low-conductance peaks and have been identified in molecular junctions with thiophene anchoring groups\cite{tiophene}.
In this way, a second surface scan was performed; this time by elongating the Au$-$H distance (see Fig. \ref{fig:PotentialEnergy}b).
A clear minimum is observed at 2.18 \AA{} when using a dispersion corrected density functional (D3 correction and BP86 functional).
The corresponding Au$-$C(9) distance is 7.49 \AA{} in this conformation.

\begin{figure}[H]
	\includegraphics[width=\linewidth]{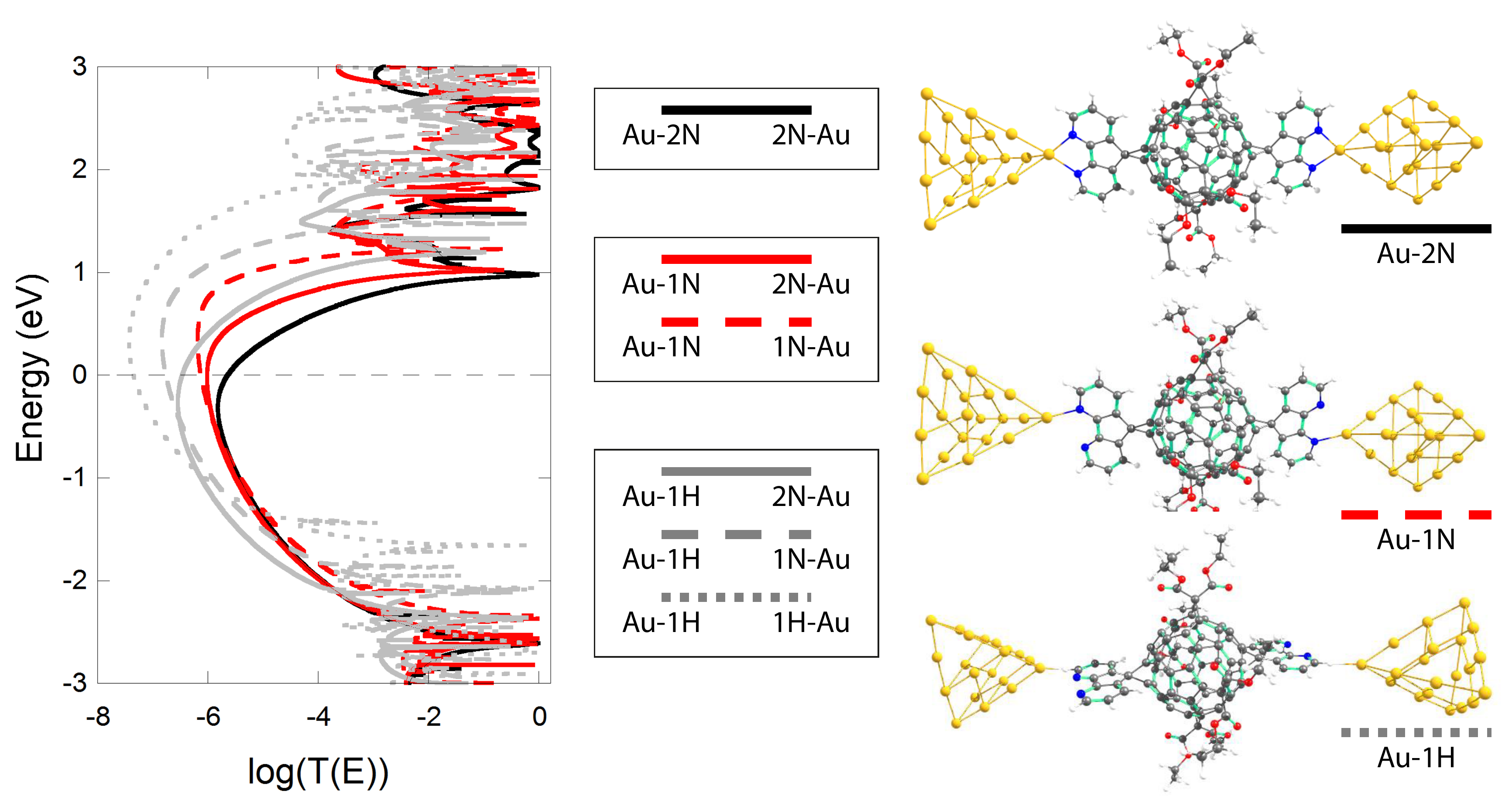}
   \caption{Transmission function for the different binding models.
   The line colour represents the number of Au$-$N bonds at one gold tip and the line style represents the binding mode at the other end.
   The solid black line indicates binding by two N atoms at each side.
   The dashed red line indicates binding by one N atom at each diazafluorene anchoring group and the grey dotted line indicates a Au$-$H van der Waals contact at each side.
   Mixed colours and line styles mean mixed binding modes.
   For instance, a dashed grey line corresponds to a Au$-$H contact on one tip (for the grey colour) and one N binding on the opposite side (dashed line).
   The Fermi level is located at 0 eV and is indicated by a horizontal dashed line.}
    \label{fig:Transmission}
\end{figure}

The three binding motifs can be combined to yield nine different anchoring patterns to the gold tips: ranging from a strongly bonded geometry where both diazafluorene groups are bonded through their two N atoms, to a weakly bonded case with two Au$-$H interactions.
Regardless of the specific binding mode, all calculations show a relatively large band gap (>3 eV), where the LUMO orbital tends to be closer to the Fermi level (between 1 and 1.5 eV above $E_\textup{F}$) and should therefore dominate electron transport.
Despite the relatively large extension of the molecule, pronounced conductance peaks can be observed.
In Fig. \ref{fig:Transmission}, broad maxima are present at $-$2.5 eV, 1 eV and 2-3 eV, where the Fermi level is set to 0 eV.
This is due to the delocalized nature of the conducting orbitals, associated with the $\pi$-systems of the fullerene and the diazafluorene moieties.
In this way, chemical modifications altering the frontier orbital energies should be efficient in tuning electron transport in this system, as these orbitals appear as efficient conduction channels due to their extensive delocalization. 

It is interesting to consider how the transmission function at the Fermi level is affected by the different coordination modes of the diazafluorene ligands (Fig. \ref{fig:Transmission}).
As expected, the model considering both diazafluorene groups coordinated by two N atoms (2$-$2 binding) presents the highest conductance (black solid line).
If one of the four N atoms is uncoordinated, one side will have two Au$-$N bonds and the opposite electrode will present only one bond (2$-$1 binding).
This situation is represented with a red solid line in Fig. \ref{fig:Transmission} and is characterized with a significantly lower conductance than the completely coordinated model.
As a result, both coordination modes should correspond to different conductance peaks (A and B).
If another Au$-$N bond is broken to yield a 1$-$1 bonding geometry (red dashed line), we observe a similar conductance to the 2$-$1 motif.
The conductance of both situations is remarkably similar and should be indistinguishable in break-junction experiments (both are assigned to peak B).
For the 1$-$1 bonding geometry, we also verified the effect of the relative orientation of the binding N atoms, calculating the `cis' like conformation (considering the red dashed line geometry as `\textit{trans}' because the Au$-$N bond is placed at opposite sides in each anchoring group).
Again, the change in conductance is negligible.

Calculations considering weak Au$-$H contacts are depicted in grey.
We observe two clear groups, with the one combining Au$-$H contacts and Au$-$N bonds at a markedly larger conductance than the geometry considering only Au$-$H interactions (dotted grey line).
To summarize, we relate peak A with the highest conductance curve in which both diazafluorene groups are coordinated by two N atoms, and peak B with the binding motifs which include only Au$-$N bonds, where at least one of the binding modes is by only one N atom (red lines).
Peak C is consistent with the presence of one Au$-$H interaction and explains the lower yield for this junction geometry, as it involves weak interactions.

The comparison of the relative values of the calculated transmission functions at the Fermi level and the experimental conductance reinforces this assignment.
In log units, the difference between the position of peak A and B ranges from 0.5 to 0.6, while the calculated values for T(E) between black and red lines ranges from 0.37 to 0.46.
In the experiment, peaks A and C are separated by 1.1-1.3 log units, comparing favourably with the difference between the black and the first group of grey curves (0.82-1.18).
The lowest curve (dashed grey line) is 1.7 log units less conductive than the most conductive mode, and is unlikely to be detected in the experiment.
Furthermore, having weak contacts at both golds tips simultaneously leads to a weak binding mode, for which it may be difficult to obtain well-defined conductance plateaus.

The fact that peak B and C each incorporate two different binding arrangements can also explain why they are generally broader than peak A, which only includes one: they comprise different configurations which have slightly different conductance values.
This can result in a broadening of the conductance peaks when enough statistics is acquired.

\section{Conclusions}
We measured C$_{60}$-daf with the MCBJ technique in two different laboratories (Delft and Santiago de Chile).
The measurements from both locations agree in identifying three most probable conductance values that we attribute to different configurations between the diazafluorene anchoring groups and the electrodes.
Through DFT calculations using the B3LYP density functional, we relate the most conductive peak A with the configurations where both Au$-$N bonds are present at each side, the middle peak B with the Au$-$N bonds where at least one of the binding modes is through only one N atom, and the lowest conductance peak C with the presence of one Au$-$H interaction.
The presence of multiple stable arrangements and the relatively low conductance values make this derivative as-is not appealing for employment in single-molecule devices.
However, these results show that this geometry is promising and, with some modifications, could open the way to further studies in which the C$_{60}$ backbone can be modified with different functional groups.

\begin{acknowledgement}

The work at TUDelft was supported by the EU through an advanced ERC grant (Mols@Mols); device fabrication was done at the Kavli Nanolab at Delft.
The work at University of Chile was supported by Fondecyt Regular Project 1161775 (M.S. and J.L.M.), Fondecyt Regular Project 1140770, EU RISE (DAFNEOX) project SEP-210165479  (D.D.) and CONICYT/Fondecyt Postdoctoral Project 3150674 (C.G.C.).
D.A. thanks CONICYT + PAI `Concurso nacional de apoyo al retorno de investigadores/as desde el extranjero, convocatoria 2014 82140014' for financial support. Powered@NLHPC: This research was partially supported by the supercomputing infrastructure of the NLHPC (ECM-02). 
L.E. thanks the National Science Foundation [grant CHE-1408865] and the PREM Program [grant DMR-1205302] as well as the Robert A. Welch Foundation [grant AH-0033] for generous financial support.
D.S. thanks Riccardo Frisenda for providing the MCBJ schematics in Fig. \ref{fig:schematics}b.
\end{acknowledgement}

This document is the unedited Author's version of a Submitted Work that was subsequently accepted for publication in Chemistry of Materials, copyright \textcopyright American Chemical Society after peer review. To access the final edited and published work see \url{http://pubs.acs.org/doi/abs/10.1021/acs.chemmater.7b02037}].

\begin{suppinfo}
Supplementary figures, analysis of additional samples and with different parameters, computational details and synthetic procedures are available in the Supporting Information.
\end{suppinfo}

\bibliography{C60-bibliography-short}

\end{document}